\documentclass[letterpaper,english,reprint,pre,longbibliography,aps,superscriptaddress]{revtex4-1}
\usepackage[T1]{fontenc}
\usepackage[latin9]{inputenc}
\usepackage{babel}
\usepackage{verbatim}
\usepackage{mathtools}
\usepackage{bm}
\usepackage{amsmath}
\usepackage{amssymb}
\usepackage{graphicx}
\usepackage{color}
\usepackage[unicode=true,
 bookmarks=true,bookmarksnumbered=false,bookmarksopen=false,
 breaklinks=false,pdfborder={0 0 1},backref=false,colorlinks=false]
 {hyperref}

\usepackage{ulem}
\usepackage{bbold}

\makeatletter

\pdfpageheight\paperheight
\pdfpagewidth\paperwidth

\newcommand{\traj}{\varphi}
\newcommand{\tfinal}{{t}_f}
\newcommand{\Ptube}{P^{\,{\varphi}}_R}
\newcommand{\aexit}{\alpha_{R}^{\traj}}

\newcommand{\xinitial}{x_0}
\newcommand{\xfinal}{x_f}

\newcommand{\xleft}{x_{\mathrm{left}}}
\newcommand{\xright}{x_{\mathrm{right}}}
\newcommand{\xtop}{x_{\mathrm{top}}}

\newcommand{\trajTwo}{\psi}

\newcommand{\aexitfree}{{\alpha}_{\mathrm{free}}}
\newcommand{\afree}{\alpha_{\mathrm{free}}}

\makeatother

\begin{document}
\title{Resolution dependence of most probable pathways with state-dependent diffusivity}
\author{Alice L.~Thorneywork}
\affiliation{Physical and Theoretical Chemistry Laboratory, University of Oxford, South Parks Rd, Oxford OX1 3QZ, United Kingdom}
\affiliation{Cavendish Laboratory, University of Cambridge, J J Thomson Ave, Cambridge CB3 0HE, United Kingdom}
\author{Jannes Gladrow}
\affiliation{Microsoft Research, Station Rd, Cambridge CB1 2FB, United Kingdom}
\author{Ulrich F.~Keyser}
\affiliation{Cavendish Laboratory, University of Cambridge, J J Thomson Ave, Cambridge CB3 0HE, United Kingdom}
\author{Michael E.~Cates}
\affiliation{Department of Applied Mathematics and Theoretical Physics, Centre for Mathematical Sciences, University of Cambridge, Wilberforce Rd, Cambridge CB3 0WA, United Kingdom}
\author{Ronojoy Adhikari}
\affiliation{Department of Applied Mathematics and Theoretical Physics, Centre for Mathematical Sciences, University of Cambridge, Wilberforce Rd, Cambridge CB3 0WA, United Kingdom}
\author{Julian Kappler}
\affiliation{Department of Applied Mathematics and Theoretical Physics, Centre for Mathematical Sciences, University of Cambridge, Wilberforce Rd, Cambridge CB3 0WA, United Kingdom}
\affiliation{Arnold Sommerfeld Center for Theoretical Physics (ASC), Department of Physics, Ludwig-Maximilians Universit\"at M\"unchen, Theresienstra{\ss}e 37, D-80333 Munich, Germany}
\date{\today}
\begin{abstract}
Recent experiments have probed the relative likelihoods of trajectories in stochastic systems by observing survival probabilities within a tube of radius $R$ in spacetime. We measure such probabilities here for a colloidal particle in a corrugated channel, corresponding to a bistable potential with state-dependent diffusivity. 
In contrast to previous findings for state-independent noise,
we find that the most probable pathway changes qualitatively as the tube radius $R$ is altered. We explain this by computing the survival probabilities predicted by overdamped Langevin dynamics. At high enough resolution (small enough $R$), survival probabilities depend solely on diffusivity variations, independent of deterministic forces;
finite $R$ corrections yield a generalization of the Onsager-Machlup action.
As corollary, ratios of survival probabilities are singular as $R \to 0$, 
but become regular, and described by the classical Onsager-Machlup action, 
only in the special case of state-independent noise.
\end{abstract}
\maketitle

In the study of rare diffusive events it is often of interest to know not only the rate at which an initial state transits to a final state but also the most likely sequence of states that connects the initial and final states \cite{dykman_optimal_1992,
luchinsky_analogue_1998,e_string_2002,
ren_transition_2005,e_transition_2005,
chan_paths_2008,fujisaki_onsagermachlup_2010,
schorlepp_gelfandyaglom_2021}. 
Care needs to be taken in the definition of the most probable path, or more generally in the quantification of relative probabilities for pairs of paths, since, apriori, single trajectories in diffusive dynamics have vanishing probability. A natural and experimentally measurable resolution consists, instead, of considering the probability of trajectories to remain within a ball of finite radius $R$ around a smooth path \cite{ventsel_small_1970,
stratonovich_probability_1971,
durr_onsager-machlup_1978,
williams_probability_1981,
horsthemke_onsager-machlup_1975,
zeitouni_onsager-machlup_1989,
ito_probabilistic_1978,
fujita_onsager-machlup_1982,
ikeda_stochastic_1989,
stratonovich_probability_1971,
horsthemke_onsager-machlup_1975,
kappler_stochastic_2020},
 which represents a tube in spacetime. The corresponding most probable tube (MPT) can, then, be defined as the path that maximises this sojourn probability between given initial and final states \cite{durr_onsager-machlup_1978}.  More generally, ratios of probabilities in tubes of equal radius $R$ can be considered, and the limit $R \rightarrow 0$ (should it exist) can be used to  define the relative probabilities of paths.  Indeed, in recent experimental work \cite{gladrow_experimental_2021}, the MPT for diffusion processes with state-independent diffusivity  was inferred from measured sojourn probabilities and
extrapolated to zero radius. 
Thus experiments can now provide a direct observational underpinning for previously theoretical constructs, such as the identification of most probable paths (MPPs) as minima of a stochastic action. Indeed, the experiments of Ref.~\cite{gladrow_experimental_2021} confirmed that, for a colloidal system with Langevin dynamics of state-independent diffusivity, the MPPs minimize the Onsager-Machlup (OM) action. This implies, and the experiments confirm, that the vanishing-radius limit of sojourn probability ratios remains well defined for any pair of tubes.

However, in many physical systems the diffusivity is state-dependent \cite{kampen_stochastic_2007,
hummer_position-dependent_2005,
sedlmeier_water_2011,
berezhkovskii_time_2011,
bo_functionals_2019,
murphy_brownian_1972,
wilemski_derivation_1976,
singh_fluctuating_2017,
li_particle_2020}. 
Examples include colloidal particles that interact hydrodynamically with other particles and container boundaries \cite{murphy_brownian_1972,
wilemski_derivation_1976,
singh_fluctuating_2017},
and Brownian particles in thermal gradients \cite{bo_functionals_2019}. 
While for these scenarios D\"urr and Bach showed mathematically that there 
exists no functional that describes MPPs \cite{durr_onsager-machlup_1978}, 
several possible stochastic actions have been proposed  \cite{stratonovich_probability_1971,
graham_path_1977,
ito_probabilistic_1978,
langouche_functional_1979,
wissel_manifolds_1979,
dekker_path_1980,
williams_probability_1981,
fujita_onsager-machlup_1982,
ikeda_stochastic_1989,
cugliandolo_building_2019}. 
However, hitherto neither the non-existence theorem nor the proposed stochastic actions
have been related to experimentally observed finite-temperature diffusive dynamics.

In this paper and the accompanying Ref.~\cite{accompanying_paper} we provide this connection,
and resolve the seeming contradiction between the D\"urr-Bach result \cite{durr_onsager-machlup_1978}
and the existence of multiplicative-noise actions  \cite{stratonovich_probability_1971,
graham_path_1977,
ito_probabilistic_1978,
langouche_functional_1979,
wissel_manifolds_1979,
dekker_path_1980,
williams_probability_1981,
fujita_onsager-machlup_1982,
ikeda_stochastic_1989,
cugliandolo_building_2019,
de_pirey_path_2023}. We consider
 the sojourn probability for diffusion processes with state-dependent diffusivity,  
 and probe the corresponding MPT experimentally for a colloidal particle in a corrugated microchannel, as a paradigmatic example of physical systems with state-dependent diffusivity. 
 We find that the MPT is determined by a radius-dependent competition between the mean diffusivity along the path and a generalized OM action, which we derive in the accompanying Ref.~\cite{accompanying_paper}. Importantly, the latter is only a subdominant contribution to the path-dependence of the sojourn probability, and becomes irrelevant in the single-path limit of asymptotically small radius. We thus demonstrate that there is no simple connection between the multiplicative-noise stochastic actions in the literature and physical MPPs, which experimentally confirms the non-existence theorem \cite{durr_onsager-machlup_1978}. More generally our results demonstrate that ratios of path probabilities of Langevin dynamics are typically singular, and are finite only when the noise is additive.

\begin{figure}[ht!]
\centering
  \includegraphics[width=0.855\columnwidth]{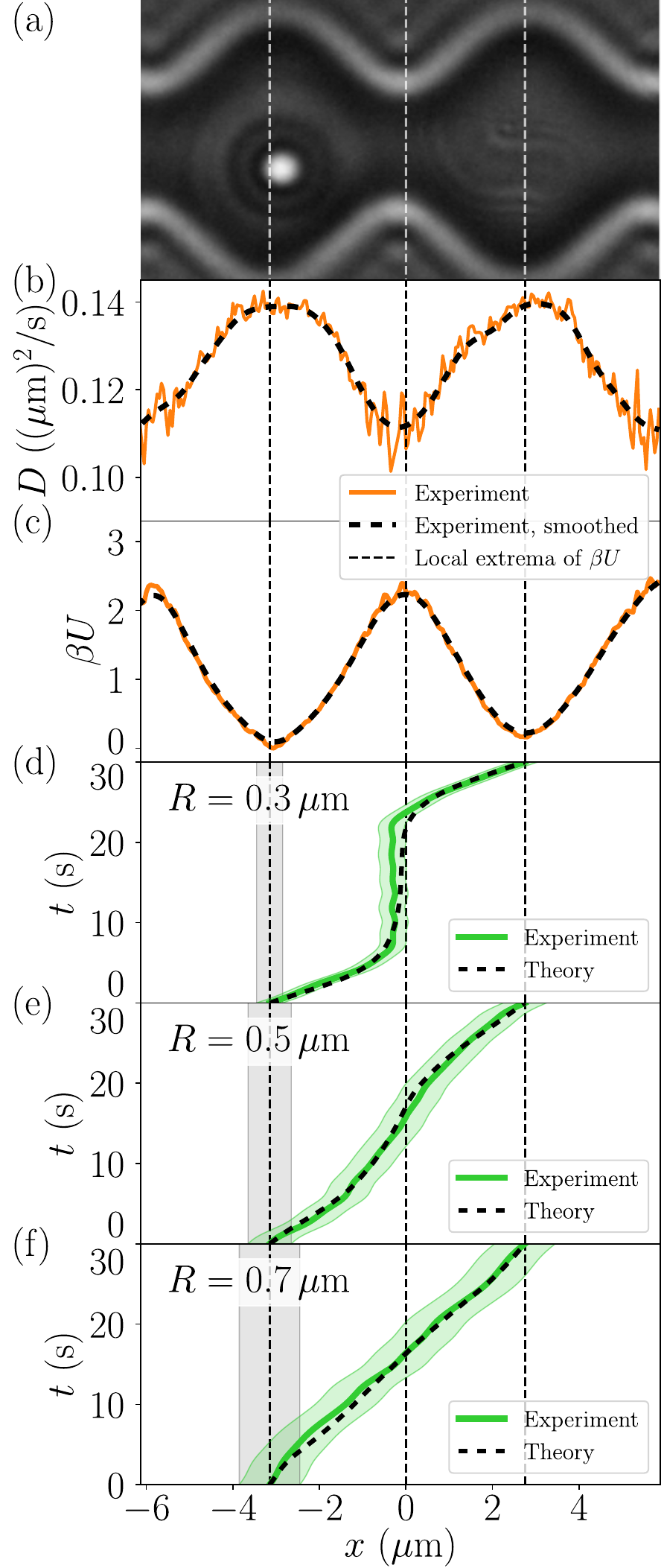} 
\caption{ \label{fig:setup} 
(a) 
Our experimental setup consists of a 
colloidal particle in a corrugated channel. 
(b) Diffusivity and (c) potential profile inferred from
 recorded time series (orange solid line);
black dashed lines are obtained by smoothing using a Hann window
average of width $1.5\,\mu$m.
(d), (e), (f)
Most probable tube (MPT) for barrier crossing for several values of $R$, inferred
 from the experimental data via the cloning algorithm from Ref.~\cite{gladrow_experimental_2021}
 (green solid lines), and corresponding finite-radius tube (green shaded region).
Black dashed lines show theoretical MPT obtained via
maximizing Eq.~\eqref{eq:most_probable_tube} using Eqs.~\eqref{eq:S_def}, 
\eqref{eq:axit_power_series};
for details see SM \cite{supplemental}.
Vertical dotted lines denote local extrema of smoothed
potential.
}
\end{figure}

\textit{Sojourn probability, tubular stochastic action, and most probable tube.}
For a smooth reference path $\traj$ and tube radius $R$, we consider the 
tubular ensemble
 comprised of all stochastic trajectories that remain within a constant distance $R$ to $\traj$
 until a final time $\tfinal$.
The probability that a stochastic trajectory
that starts within the tube
is still in the tubular ensemble at time $\tfinal$ is called the 
 sojourn probability
$\Ptube(\tfinal)$.

This probability is equivalently described by
 the instantaneous exit rate $\aexit$ at which 
stochastic trajectories leave the tube for the first time via \cite{kappler_stochastic_2020,gladrow_experimental_2021}
\begin{align}
\Ptube(\tfinal)&=
\label{eq:exit_rate_definition}
\exp\left[-\int_{0}^{\tfinal}\mathrm{d}t~\aexit(t)\right].
\end{align}
We define the negative exponent in this expression as the finite-radius stochastic action functional
\begin{align}
\label{eq:S_def} 
  S_R[\traj] \equiv -\ln(\Ptube) &= \int_{0}^{\tfinal}\mathrm{d}t~\aexit(t) ,
\end{align}
which fully describes the sojourn probability (modulo dependence on initial conditions,
which affect $\aexit$ for a short initial duration).
From Eq.~\eqref{eq:S_def} it is apparent that $\aexit$ can be interpreted as a Lagrangian
that quantifies sojourn probabilities.

The most probable tube (MPT)
for given initial and final positions $x_0$, $x_f$, total duration $\tfinal$,
and tube radius $R$ is defined as
\begin{align}
\label{eq:most_probable_tube}
\traj^*_R &\equiv \underset{\traj}{\mathrm{argmin}} \,S_R[\traj],
\end{align}
where we minimize over paths $\traj$ with
$\traj(0) \equiv \xinitial$, $\traj(\tfinal) = \xfinal$. 
According to Eqs.~\eqref{eq:exit_rate_definition}, \eqref{eq:S_def}, 
the MPT center
$\traj^*_R$ maximizes the sojourn probability $\Ptube$;
the corresponding tube therefore represents the pathway from $\xinitial$ to $\xfinal$
along which a stochastic trajectory is most likely to remain within a distance
$R$ from the reference path.

\begin{figure*}[ht]
\centering \includegraphics[width=1\textwidth]{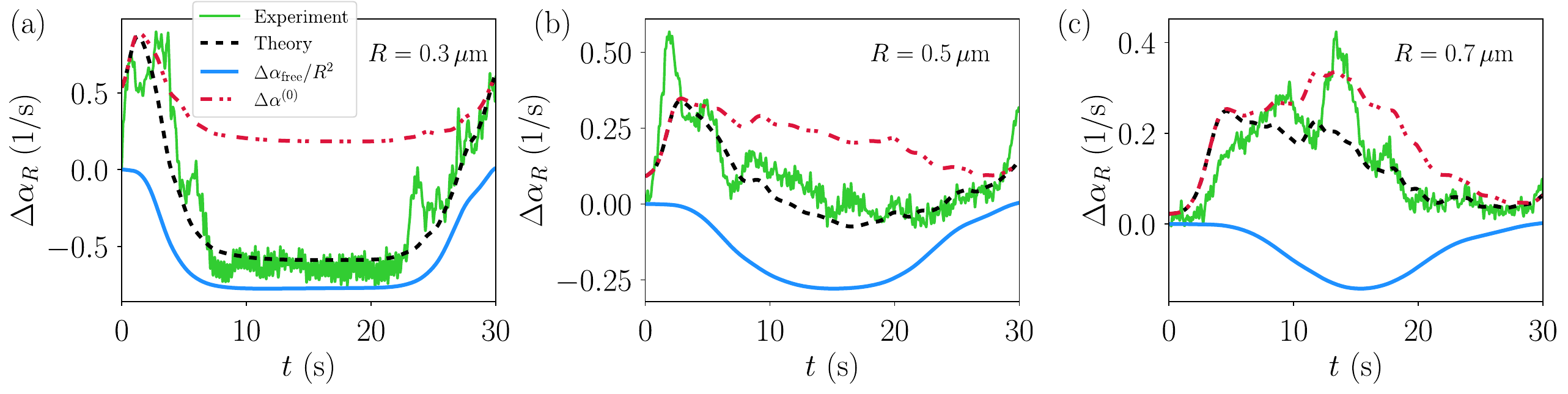} 
\caption{ \label{fig:exit_rate_differences} 
Exit-rate difference Eq.~\eqref{eq:Delta_alpha}
for (a) $R = 0.3\,\mu$m, (b) $R = 0.5\,\mu$m, (c) $R = 0.7\,\mu$m.
Green solid line depicts experimental results obtained by directly measuring
 sojourn probabilities \cite{gladrow_experimental_2021,supplemental};
black dashed line shows analytical results obtained from Eq.~\eqref{eq:axit_power_series} 
using smoothed diffusivity and potential profiles from Fig.~\ref{fig:setup} (b), (c).
Differences 
in individual terms
of Eq.~\eqref{eq:axit_power_series}
are shown as blue solid and red dash-dotted lines.
}
\end{figure*}

\textit{Experimental setup.}
In our experiments we consider a colloidal particle moving in a corrugated
 channel, c.f.~Fig.~\ref{fig:setup} (a). The colloidal system consists of carboxylate
  functionalised polystyrene particles with diameter $\sigma=1.2\,\mu m$ in 
  5mM KCl solution. 
  The microfluidic channel has periodic oscillations in width perpendicular to the
   channel axis, which vary from approximately $1.6\sigma$ to $4.8\sigma$ at the narrowest
    and widest sections of the channel respectively. 
    The height of the channel (perpendicular to the imaging plane) is on the order 
    of the particle diameter and does not vary throughout the design. 
    This results in quasi-two dimensional diffusion of the particle through 
    the channel and approximately flat probability distribution of the particle 
    position in two dimensions. 
    Trajectories are acquired using an automated data acquisition protocol 
    implemented using Holographic Optical Tweezers, as described previously 
    \cite{thorneywork_direct_2020,gladrow_experimental_2021}.

We record several time series with an observation timestep $\Delta t = 2\,$ms, 
and retain only the $x$-component (along the channel axis) to accumulate 
approximately 6 hours of sample trajectories. 
By considering only the $x$-component of the trajectory, 
we project the true two-dimensional diffusion onto a
 one-dimensional description \cite{kalinay_projection_2005,
dorfman_assessing_2014}.
  The varying channel width gives rise to
   an effective \textcolor{black}{entropic} potential, 
   with constrictions acting as entropic barriers. 
   The changing confinement of the particle along the $x$ direction 
   also leads to variation in the \textcolor{black}{effective 1D} diffusion coefficient of the particle, 
   which is strongly affected by the variation in hydrodynamic drag 
   arising from differing proximity to the channel walls \cite{yang_hydrodynamic_2017}. 
   As detailed in the SM \cite{supplemental}, 
   we 
estimate the diffusivity and potential profile 
along the $x$-direction
from our data;
   we show the resulting profiles
    in Fig.~\ref{fig:setup} (b), (c).
Figure \ref{fig:setup} (a), (b), (c)
   shows that in regions where the channel has its maximal width, 
   the diffusivity is maximal and the potential landscape features
   local minima
    ($\xleft \approx -3.2\,\mu$m, 
$\xright \approx 2.8\,\mu$m).
On the other hand, at locations where the channel is most narrow, the diffusivity shows local minima and the potential landscape shows local maxima ($\xtop =0$).

\textit{Experimental results.}
We consider the MPT
for a barrier crossing from $\xinitial = \xleft$ to
 $\xfinal = \xright$, and with $\tfinal = 30\,$s.
We infer the MPTs from experimental time series using the cloning 
algorithm from Ref.~\cite{gladrow_experimental_2021} (for a python 
implementation, see Ref.~\cite{module_cloning_algorithm}).
In Fig.~\ref{fig:setup} %(d), (e), (f) 
we
show the resulting experimental
 MPTs 
for (d) $R = 0.3\,\mu$m, (e) $R = 0.5\,\mu$m, and
(f) $R = 0.7\,\mu$m.
From Fig.~\ref{fig:setup} (d) we observe that for $R = 0.3\,\mu$m
the MPT rests close to the 
barrier top for about half of the total path duration.
This is qualitatively different from the paths in Fig.~\ref{fig:setup} (e), (f)
which cross over the barrier without lingering there.

For the three MPTs, we in Fig.~\ref{fig:exit_rate_differences} 
plot the experimentally measured exit-rate difference
\begin{align}
\label{eq:Delta_alpha}
\Delta \alpha_R(t) &\equiv \alpha_R^{\traj^*}(t) - \alpha_R^{\trajTwo}(t),
\end{align}
between the MPT $\traj^*$ and a second path $\trajTwo$, 
which rests at the left local minimum $\xleft$, as illustrated by gray shaded areas in 
Fig.~\ref{fig:setup} (d-f).
Since $\trajTwo$, drift, and diffusivity are independent of time, 
so is $\alpha_R^{\trajTwo}$.
Any time-dependence in $\Delta \alpha_R$
thus originates from
$\alpha_R^{\traj^*}$
\textcolor{black}{(in the SM \cite{supplemental} we discuss the
 advantage of considering  Eq.~\eqref{eq:Delta_alpha} over
$\alpha_R^{\traj^*}$ alone)}.
In Fig.~\ref{fig:exit_rate_differences} (a) we see that for $R = 0.3\,\mu$m the
experimental exit-rate difference is negative for times during which the MPT rests close 
to the barrier top. As such, 
according to Eq.~\eqref{eq:Delta_alpha}
the exit rate at the minimum $\xleft$ is larger
than at the barrier top.
A tube lingering at the potential barrier top is hence more probable
than a tube lingering at the potential minimum; we discuss this at first sight
counterintuitive result further below.
For 
$R = 0.5\,\mu$m, $R = 0.7\,\mu$m, 
 the exit-rate differences in Fig.~\ref{fig:exit_rate_differences} (b), (c) 
 are predominantly
positive, so that the exit rate along the barrier-crossing tube is larger
than at the left minimum.

\textit{Theoretical results.}
To understand the observed crossover from barrier-preferring
 tubes at small radius
 to well-preferring tubes at larger radius, 
we consider a stochastic process $X_t$ described by
 the It\^{o} equation
\begin{align}
\label{eq:Langevin_equation}
dX_t &= a(X_t)dt + \sqrt{ 2 D(X_t)}dW_t,
\end{align}
with drift $a(x)$,
state-dependent diffusivity $D(x)$, and
 increment of the Wiener process $dW_t$.
We note that $a(x)$ is closely related to the spatial derivative of the
 potential $U$ from Fig.~\ref{fig:setup} (c), 
see SM \cite{supplemental} for details.
For the dynamics defined by Eq.~\eqref{eq:Langevin_equation}, we
in Ref.~\cite{accompanying_paper}
 calculate $\aexit$
 as power series in the tube radius $R$,
resulting in 
\begin{align}
\aexit(t) & =
\frac{\alpha_{\mathrm{free}}^{\traj}(t)}{R^2} +
\alpha^{\traj,(0)}(t)
\label{eq:axit_power_series}  
+\mathcal{O}(R^{2}),
\end{align}
with
\begin{align}
\label{eq:afree_1D}
\frac{\aexitfree^{\traj} (t)}{R^2}&=   \frac{\pi^2}{4}\frac{D(\traj(t))}{R^2}.
\end{align}
Equation \eqref{eq:afree_1D} is the steady-state absorbing-boundary exit rate 
on a domain $[-R,R]$
for diffusive dynamics with a state-independent diffusivity
equal to $D(\traj(t))$ and vanishing drift
\cite{kappler_stochastic_2020,accompanying_paper};
we hence call $\aexitfree^{\traj}/R^2$ the free-diffusion exit rate.

For small radius
Eq.~\eqref{eq:axit_power_series}
is dominated by
 the free-diffusion contribution Eq.~\eqref{eq:afree_1D},
which depends on $D(\traj(t))$.
For state-dependent diffusivity the
 exit rate Eq.~\eqref{eq:axit_power_series}
 is thus path-dependent to leading order in the limit $R \rightarrow 0$.
This is the 
 key conceptual difference between 
 sojourn probabilities for
  multiplicative and additive \cite{kappler_stochastic_2020} noise (where 
  $D$ is constant).

The first correction to free diffusion in Eq.~\eqref{eq:axit_power_series} 
is  \cite{accompanying_paper}
\begin{align}
\alpha^{\traj,(0)} &= 
 \frac{1}{4D} \left( \dot{\traj} - a  \right)^2
+\frac{1}{2}  \partial_x a
 + \frac{1}{2} \frac{ \partial_x D}{D} \left( \dot{\traj} -a  \right)
\label{eq:Lagrangian_1D}
\\ &\quad \qquad \nonumber
+c_1 \frac{ \left(\partial_x D\right)^2}{D}
+c_2 \partial_x^2 D,
\end{align}
where drift, diffusivity, and their derivatives, are evaluated along $\traj$,
and $c_1 = - ( \pi^2 -3)/16$, $c_2 =  ( \pi^2 - 6 )/24$.
Equation \eqref{eq:Lagrangian_1D} is a
 multiplicative-noise generalization of the OM action \cite{stratonovich_probability_1971,
durr_onsager-machlup_1978,ito_probabilistic_1978,
williams_probability_1981,ikeda_stochastic_1989,
kappler_stochastic_2020}, 
to which it reduces for state-independent diffusivity. 
The appeal of our Eq.~\eqref{eq:Lagrangian_1D},
which is different from the 
multiplicative-noise Lagrangians 
in the literature \cite{stratonovich_probability_1971,
durr_onsager-machlup_1978,
cugliandolo_building_2019,stratonovich_probability_1971,horsthemke_onsager-machlup_1975,
fujita_onsager-machlup_1982,ikeda_stochastic_1989,
ito_probabilistic_1978,williams_probability_1981},
 is its direct
relation to the physical observable Eq.~\eqref{eq:axit_power_series}.

In Fig.~\ref{fig:setup} (d), (e), (f) and Fig.~\ref{fig:exit_rate_differences}
we show theoretical MPTs and corresponding 
exit-rate differences.
For 
all values of $R$ the theoretical and experimental MPTs in Fig.~\ref{fig:setup}
agree well. 
While the corresponding theoretical exit rates (black dashed lines 
in Fig.~\ref{fig:exit_rate_differences})
do not reproduce all small-scale features of the experimental exit rates,
they do capture the overall behavior of the latter.
In the SM \cite{supplemental} we furthermore 
evaluate Eq.~\eqref{eq:axit_power_series} 
on the experimental MPT centers and find very good agreement with 
the corresponding experimental exit rates.

\begin{figure}[ht!]
\centering \includegraphics[width=1\columnwidth]{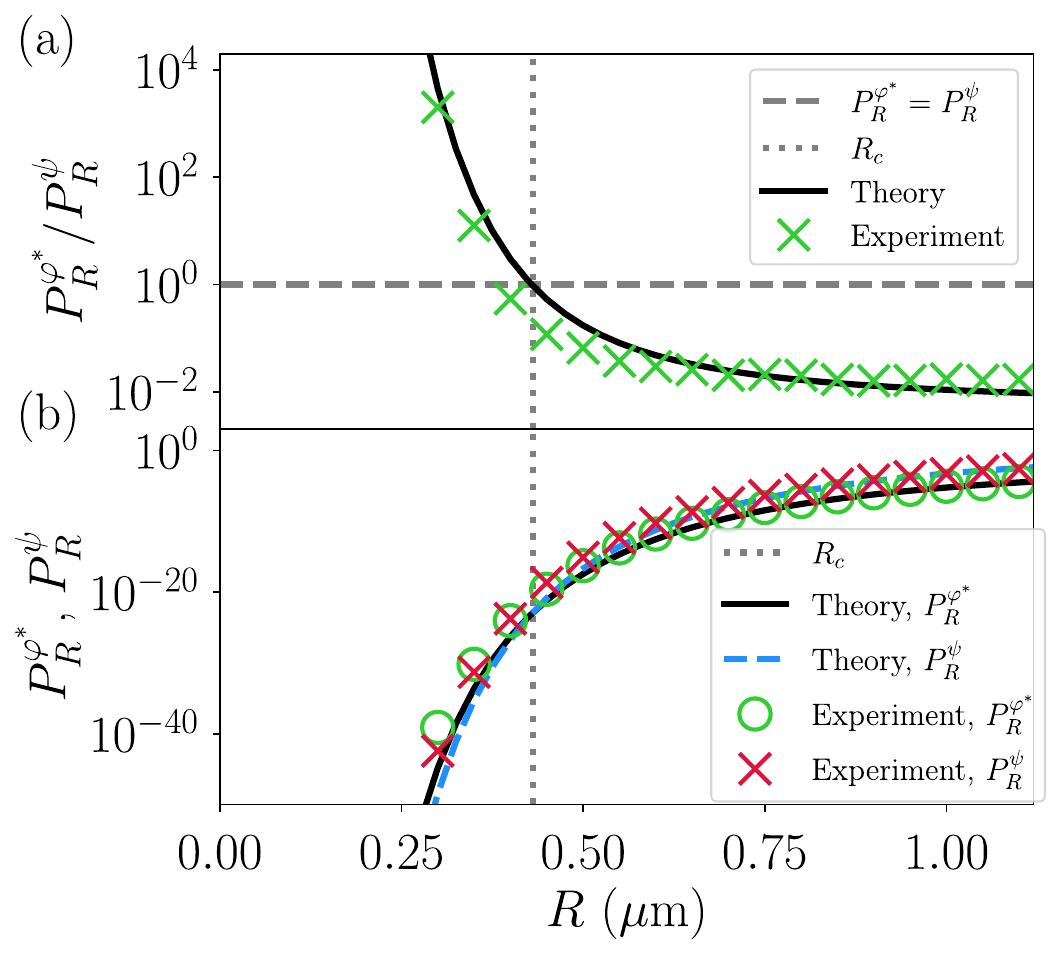} 
\caption{ \label{fig:sojourn} 
(a) Ratios of sojourn probabilities $P^{\traj^*}_R/P^{\trajTwo}_R$ for 
MPT $\traj^*$ and path $\trajTwo$ that rests at $\xleft$.
Green crosses show experimental results, obtained from measured exit-rate
differences  $\Delta \alpha_R$
via numerical evaluation of
${P_R^{\traj^*}}/{P_R^{\trajTwo}} = \exp[ - \int_0^{\tfinal}dt\,\Delta \alpha_R(t)]$.
Black curves are obtained using the same formula but with
 perturbative exit rate Eq.~\eqref{eq:axit_power_series},
and MPTs obtained by minimizing the corresponding theoretical finite-radius action.
Horizontal dotted line denotes value ${P_R^{\traj^*}}/{P_R^{\trajTwo}} = 1$.
(b)
Green and red symbols denote experimental sojourn probabilities 
$P^{\traj^*}_R$, $P^{\trajTwo}_R$ with $\traj^*$, $\trajTwo$ as in (a).
Black and orange curves represent the corresponding theoretical sojourn probabilities,
evaluated as in (a).
Vertical dotted lines in (a), (b) denote the crossover radius $R_c \approx 0.44\,\mu$m
where $P^{\traj^*}_R =P^{\trajTwo}_R$ (based on
theoretical sojourn probabilities).
}
\end{figure}

In Fig.~\ref{fig:exit_rate_differences} we
 decompose  the theoretical exit-rate difference Eq.~\eqref{eq:Delta_alpha}
  into 
$\Delta \afree /R^2 \equiv 
(\afree^{\traj} - \afree^{\trajTwo})/R^2$ 
and 
$\Delta \alpha^{(0)} \equiv
 \alpha^{\traj,(0)} -  \alpha^{\trajTwo,(0)}$. 
 Subplot (a) shows 
  that for $R = 0.3\,\mu$m the exit rate is dominated by the free-diffusion contribution
 $\Delta \afree /R^2$,
which explains why the MPT from Fig.~\ref{fig:setup} (d)
rests at the barrier top:
since Eq.~\eqref{eq:afree_1D} is proportional to the diffusivity, 
the MPT seeks out spatial regions where
the diffusivity is minimized, which for our system 
 is close to $\xtop$, c.f.~Fig.~\ref{fig:setup} (b).
On the other hand, for $R = 0.7\,\mu$m
 we see in Fig.~\ref{fig:exit_rate_differences} (c) 
 that 
 $\Delta \alpha_R$
 is dominated by 
 $\Delta \alpha^{(0)}$, so that the MPT is predominantly determined by
 Eq.~\eqref{eq:Lagrangian_1D} as opposed to Eq.~\eqref{eq:afree_1D}.
 In summary, Figs.~\ref{fig:setup}, \ref{fig:exit_rate_differences} show that
for $R \lesssim 0.7\,\mu$m the theoretical exit rate
Eq.~\eqref{eq:axit_power_series} describes the experimental results well, 
and explains  the qualitative change
 in the MPT 
 observed in Fig.~\ref{fig:setup} (d), (e), (f)
as a radius-dependent competition between the free-diffusion
 term Eq.~\eqref{eq:afree_1D} and the next-order correction Eq.~\eqref{eq:Lagrangian_1D}
 in the perturbative exit rate Eq.~\eqref{eq:axit_power_series}.

\textit{Radius-dependence of sojourn probability.}
According to Eq.~\eqref{eq:exit_rate_definition}, the difference 
Eq.~\eqref{eq:Delta_alpha}
quantifies the ratio of sojourn probabilities
${P_R^{\traj^*}}/{P_R^{\trajTwo}} = \exp[ - \int_0^{\tfinal}dt\,\Delta \alpha_R(t)]$,
which we show in Fig.~\ref{fig:sojourn} (a) 
as function of $R$.
For $R = 0.3\,\mu$m the MPT is three orders of magnitude more
probable than the constant tube resting at the potential minimum $\xleft$,
which in Fig.~\ref{fig:exit_rate_differences} (a)  we explained by 
the small-radius dominance of  
$\Delta \afree /R^2$
in Eq.~\eqref{eq:axit_power_series},
together with the fact that 
 $D(\xleft) > D(\xtop)$.
The  diverging limit $R \rightarrow 0$ in Fig.~\ref{fig:sojourn} (a) 
originates from the non-zero term $\Delta \afree /R^2$,
and illustrates that 
ratios of path probabilities are in general ill-defined for state-dependent diffusivity.
With increasing radius the free-diffusion contribution
in Eq.~\eqref{eq:axit_power_series} becomes less relevant, and we in
Fig.~\ref{fig:sojourn} (a) see that 
 at  $R_c \approx 0.44\,\mu$m
the barrier-crossing MPT becomes less
probable than the tube that rests at $\xleft$
\textcolor{black}{(in the SM we derive an analytical estimate
for $R_c$ 
\cite{supplemental}).}

In Fig.~\ref{fig:sojourn} (b) we show  
$P_R^{\traj^*}$, $P_R^{\trajTwo}$ as function of $R$.
For $R = 0.3\,\mu$m the sojourn probabilities are of the order of
$10^{-40}$.
This emphasizes how rare it is for a stochastic realization at finite temperature 
to follow a given path closely, and demonstrates the
 capabilities of our
cloning algorithm for probing rare events from data \cite{gladrow_experimental_2021,module_cloning_algorithm}.
For small radii $R \lesssim 0.7\,\mu$m the sojourn probabilities increase steeply with $R$;
 as the free-diffusion term in Eq.~\eqref{eq:axit_power_series} becomes less dominant,
 the slope decreases and the sojourn probabilities
approach their asymptotic limit $P \rightarrow 1$ (for $R \rightarrow \infty$).

\textit{Conclusions.} 
For a colloidal particle in a corrugated microchannel,
we
from experimental data
 infer the most probable tube (MPT) for a barrier crossing
 with a finite tolerance $R$ for digression from the tube center.

To obtain our results we use a
 cloning algorithm \cite{gladrow_experimental_2021,module_cloning_algorithm}, 
 which allows us to probe probability distributions on
the  infinite-dimensional space of all realizations of diffusive dynamics
from observed time series.

We find that the MPT is determined by a scale-dependent competition between
stochastic and deterministic force, made explicit via the tube radius $R$
in Eq.~\eqref{eq:axit_power_series}.
\textcolor{black}{Intuitively, 
diffusive stochastic dynamics is for asymptotically small time- and length-scales 
 dominated by random forces, and independent of the deterministic drift.}
\textcolor{black}{Tubes of asymptotically small radius probe precisely 
this small-scale behavior \cite{kappler_stochastic_2020,gladrow_experimental_2021,
accompanying_paper}.}
The small-radius limiting MPT therefore first and foremost minimizes the diffusivity along the path. Unless this diffusivity is constant, the MPT minimizes a potential-dependent stochastic action 
only for sufficiently large tube radius $R$. 
We have derived for such cases an action that generalizes the Onsager-Machlup one. 
While this has a similar form to some previous proposals in the literature  \cite{stratonovich_probability_1971,
durr_onsager-machlup_1978,
cugliandolo_building_2019,stratonovich_probability_1971,horsthemke_onsager-machlup_1975,
fujita_onsager-machlup_1982,ikeda_stochastic_1989,
ito_probabilistic_1978,williams_probability_1981}, it differs crucially from 
these in relating experimentally observable quantities, without prior knowledge of the state-dependent diffusivity itself. None of these actions can describe ratios of finite-temperature path probabilities in the experimentally transparent limit where the tube radius $R\to 0$ (while remaining constant along the path). Instead, as we have shown,
this limit is in general singular, 
with a path that experiences an on average smaller diffusivity 
being infinitely more probable as compared to a path with on average larger diffusivity.

One consequence of this is that the small-radius MPT can 
prefer lingering at potential maxima over minima, c.f.~Fig.~\ref{fig:setup} (d).
This challenges the classical view of potential minima as local points of
 stability, 
 and
  demonstrates that experimentally observable characteristics 
of multiplicative noise
systems can be radically different  from their additive-noise counterparts.
Furthermore, our results suggest that for understanding transition pathways in multiplicative-noise systems, 
focus should be moved from individual paths
towards finite-radius tubes, which describe sets of finite probability on the space of all paths.

 \begin{acknowledgments}

Work funded in part by the European Research Council under the Horizon 2020 Programme, ERC grant agreement number 740269, and by the Royal Society through grant RP1700.
A.~L.~T.~thanks her Royal Society University Research Fellowship.
J.~G.~and U.~F.~K.~were supported by the European Union's Horizon 2020 research and innovation program under European Training Network (ETN) Grant No. 674979-NANOTRANS. 
U.~F.~K.~acknowledges funding from an European Research Council Consolidator Grant (DesignerPores 647144).
J.~K.~acknowledges funding from the European Union's Horizon 2020 research and innovation programme under the Marie Sk{\l}odowska-Curie grant agreement No 101068745.

\end{acknowledgments}


\begin{thebibliography}{41}
\bibitem [{\citenamefont {Kappler}\ \emph {et~al.}()\citenamefont {Kappler},
  \citenamefont {Cates},\ and\ \citenamefont {Adhikari}}]{accompanying_paper}%
  \BibitemOpen
  \bibfield  {author} {\bibinfo {author} {\bibfnamefont {Julian}\ \bibnamefont
  {Kappler}}, \bibinfo {author} {\bibfnamefont {Michael~E.}\ \bibnamefont
  {Cates}}, \ and\ \bibinfo {author} {\bibfnamefont {Ronojoy}\ \bibnamefont
  {Adhikari}},\ }\bibfield  {title} {\enquote {\bibinfo {title} {Sojourn
  probabilities in tubes and pathwise irreversibility for it{\^o} processes},}\
  }\href@noop {} {\ }\BibitemShut {NoStop}%
\end{thebibliography}
\end{document}